\documentclass[aip,reprint]{revtex4-1}

\usepackage{graphicx}

\usepackage{draftwatermark}
\SetWatermarkText{\shortstack{
    The following article has been accepted by Applied Physics Letters.\\
    After it is published, it will be found at \color{blue}{https://aip.scitation.org/journal/apl}
}}
\SetWatermarkAngle{0}
\SetWatermarkColor{red}
\SetWatermarkLightness{0}
\SetWatermarkVerCenter{0.5in}
\SetWatermarkFontSize{7pt}

\draft 

\usepackage{mathtools}
\DeclarePairedDelimiter{\abs}{\lvert}{\rvert}

\usepackage{physics}

\newlength{\actualcolumn}
\newlength{\actualcolumnsep}

\begin{document}


\title{Wurtzite Phonons and the Mobility of a GaN/AlN 2D Hole Gas} 
\author{Samuel James Bader}
\email[]{sjb353@cornell.edu}
\homepage[]{http://sambader.net}
\affiliation{Cornell University School of Applied \& Engineering Physics, Ithaca, NY USA 14853}
\author{Reet Chaudhuri}
\affiliation{Cornell University School of Electrical \& Computer Engineering, Ithaca, NY USA 14853}

\author{Martin Schubert}
\affiliation{X Development LLC, 100 Mayfield Ave, Mountain View, CA USA 94043}
\author{Han Wui Then}
\affiliation{Intel Corporation, 2501 NE Century Blvd, Hillsboro, OR USA  97124}

\author{Huili Grace Xing}
\affiliation{Cornell University School of Electrical \& Computer Engineering, Ithaca, NY USA 14853}
\affiliation{Cornell University Dept. of Materials Science \& Engineering, Ithaca, NY USA 14853}
\affiliation{Kavli Institute at Cornell, Ithaca, NY USA 14853}
\author{Debdeep Jena}
\affiliation{Cornell University School of Electrical \& Computer Engineering, Ithaca, NY USA 14853}
\affiliation{Cornell University Dept. of Materials Science \& Engineering, Ithaca, NY USA 14853}
\affiliation{Kavli Institute at Cornell, Ithaca, NY USA 14853}
\date{\today}
\begin{abstract}
  To make complementary GaN electronics more than a pipe dream, it is essential to understand the low mobility of 2D hole gases in III-Nitride heterostructures.  This work derives both the acoustic and optical phonon spectra present in one of the most prominent p-channel heterostructures (the all-binary GaN/AlN stack) and computes the interactions of these spectra with the 2D hole gas, capturing the temperature dependence of its intrinsic mobility.  Finally, the effects of strain on the electronic structure of the confined 2D hole gas are examined and a means is proposed to engineer the strain to improve the 2D hole mobility for enhanced p-channel device performance, with the goal of enabling wide-bandgap CMOS.
\end{abstract}
\maketitle
\setlength{\actualcolumn}{\columnwidth}
\setlength{\actualcolumnsep}{\columnsep}

Decades after the celebrated invention of Mg p-doping \cite{Shuji1992} in Gallium Nitride (GaN) and the subsequent development of GaN-based LEDs, the manipulation of holes in GaN remains a fundamental challenge.  Consequently, despite the expected dominance of GaN High Electron Mobility Transistors (HEMTs) in the coming generation of power electronics \cite{Huang2017} and communications systems\cite{Yuk2017}, there is no complementary p-channel device which can be readily integrated.  This incompleteness restricts the possible circuit topologies and system designs acheivable in GaN electronics, but arises quite straightforwardly from the physics of the GaN valence band.  These bands, both heavy and deep in energy, have proven difficult to contact with typical metal workfunctions \cite{Song2010}, difficult to dope with high efficiency \cite{Kozodoy2000}, and difficult to flow current through with high conductivity.  Nonetheless, the commercial interest in generating complementary GaN-based circuits \cite{Chu2016} and scientific interest in studying highly-degenerate hole physics \cite{Chaudhuri2018} have prompted great recent progress in p-channel devices \cite{BaderEDL2018}.

\begin{figure}[t]
  \centering
  \includegraphics[width=\columnwidth]{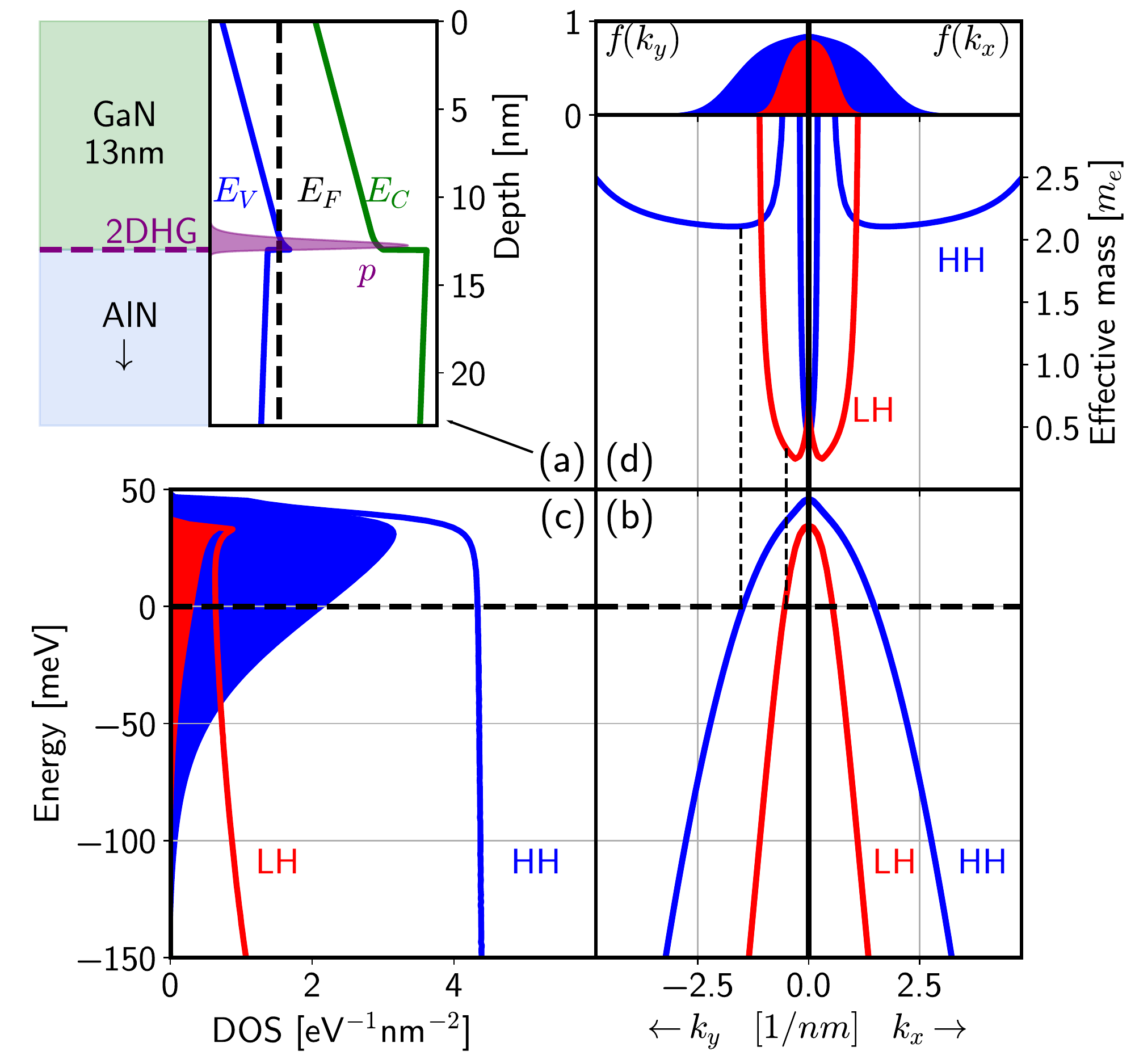}
  \caption{(a) A 13nm GaN layer on top of a thick AlN buffer induces a two-dimensional hole gas of density $\sim 4.4\times10^{13}/\mathrm{cm}^2$, which is confined at the interface by strong polarization fields. (b) The relevant bands are the first spin-degenerate HH subband-pair (blue) and LH subband-pair (red). (c) The DOS of the two bands is indicated by solid lines, and the occupation of the bands by filled shapes.  (d) The effective masses versus position in k-space, capped by the occupation probability as a function of k. The energy axes of (b) and (c) align, and the $k$-axes of (b) and (d) align. Dashed lines guide the eyes to the effective masses near the Fermi energy. }
  \label{fig:2DHG}
\end{figure}
\begin{figure*}[t]
  \centering
  \begin{minipage}[t]{\actualcolumn}
    \centering
    \includegraphics[width=\actualcolumn]{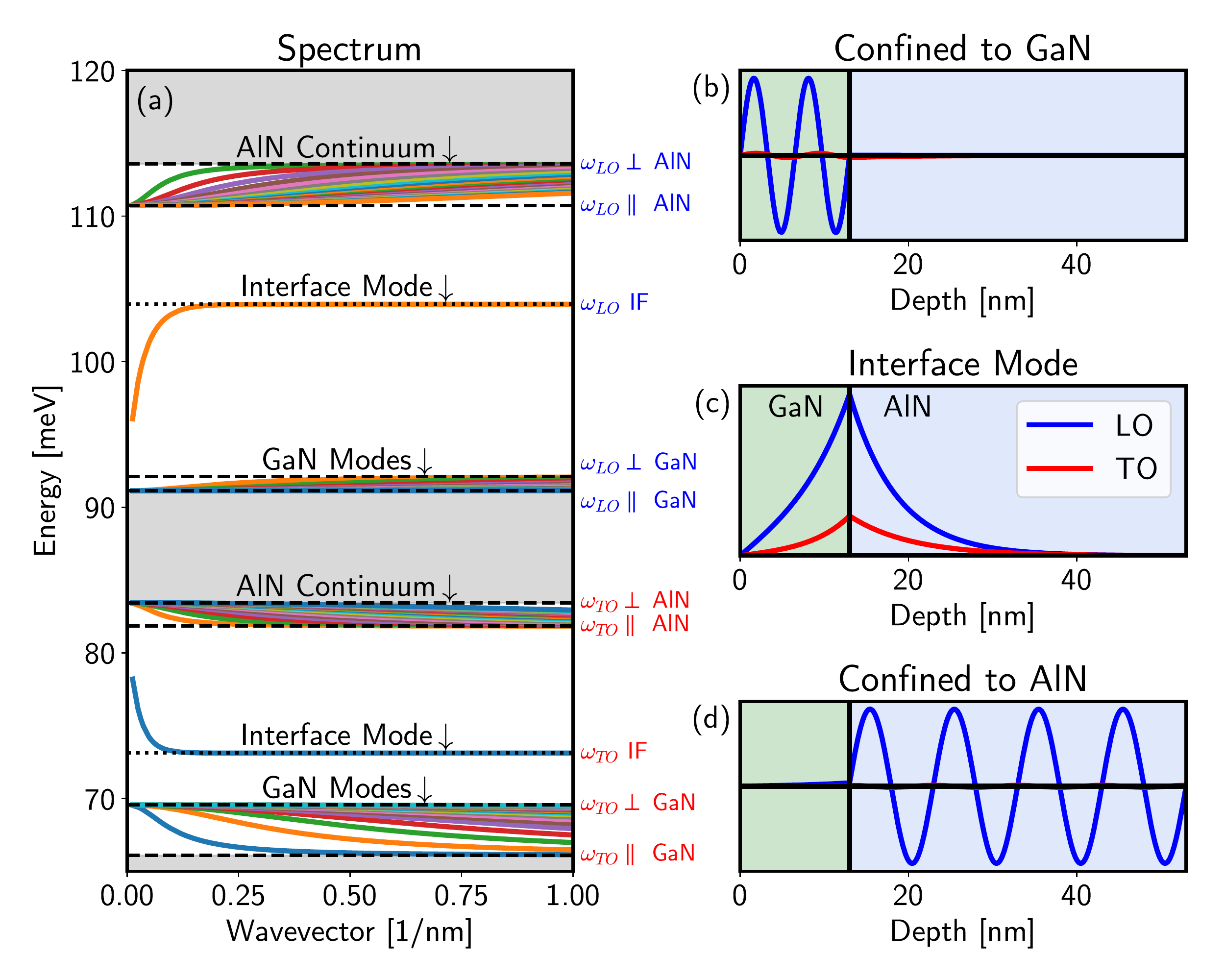}
    \caption{Polar Optical Phonons. (a) Spectrum of the various extra-ordinary POP modes, with the character of each mode labelled.  The bottom three bands (within red frequency labels) are predominantly transverse (TO), and the top three bands (within blue frequency labels) are predominantly longitudinal (LO). Examples of (b) modes confined to GaN, (c) to the interface and (d) to AlN.  For each, the potential of both possible polarizations is shown.  As expected, the TO phonons contribute weaker potential.}
  \label{fig:pop}
  \end{minipage}%
  \hspace{\actualcolumnsep}%
  \begin{minipage}[t]{\actualcolumn}
    \centering
    \includegraphics[width=\actualcolumn]{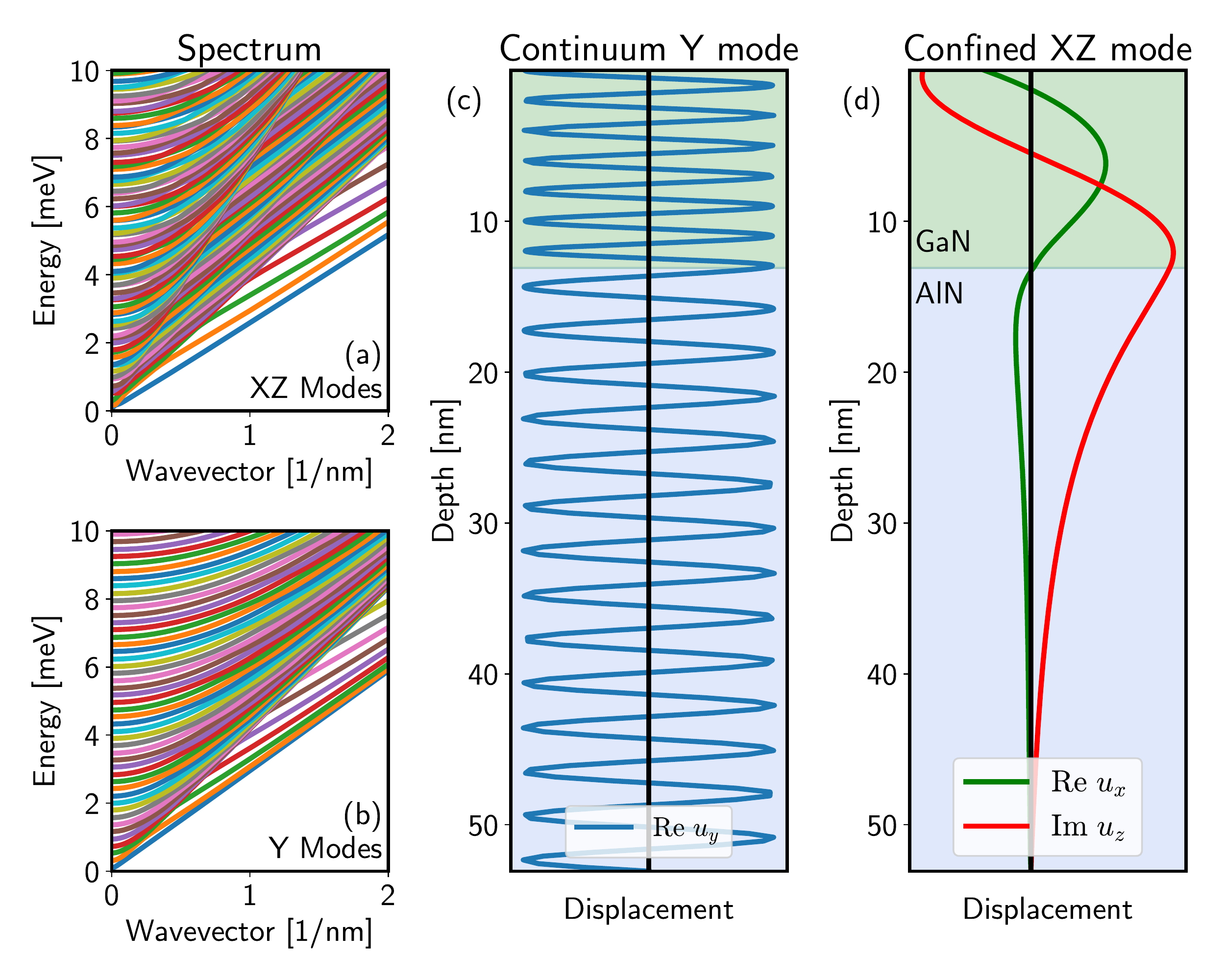}
    \caption{Acoustic Phonons. Spectrum of the (a) XZ polarized and (b) Y polarized modes.  In each there is a characteristic velocity dividing discrete modes and continuum modes: $\sqrt{C_{44}/\rho}$ and $\sqrt{(C_{11}-C_{12})/2\rho}$, respectively. 
    (c) An example of a continuum Y mode: oscillations are more concentrated in the GaN where the higher density leads to lower sound velocity. The lower sound velocity of GaN relative to AlN also allows for confinement, such as in (d), which depicts a confined XZ mode.}
  \label{fig:acoustic}
  \end{minipage}
\end{figure*}

Of the varied structures \cite{Shatalov2002, Zimmermann2004, Li2013, Hahn2013, Chu2016, Zhang2016, Nomoto2017, Nakajima2018, Krishna2019} which have been proposed as a platform for p-channel III-Nitride electronics, the single GaN/AlN heterojunction field-effect transistor has received recent attention for its high sheet conductance \cite{Chaudhuri2018} and excellent device performance \cite{BaderEDL2018}.  In this structure, depicted in Fig \ref{fig:2DHG}(a), the all-binary materials provide a straightforwardly repeatable growth with no possible parasitic electron channels, a tremendous hole-inducing polarization-charge for low sheet-resistance, and maximal bandgaps for extreme voltage-handling capability.  Given the recent reports of temperature-dependent transport studies in this heterostructure \cite{Chaudhuri2018}, and recent first-principles suggestions of possible enhancements to hole mobility in bulk p-GaN \cite{Ponce2019}, this work presents a model to explain the measured mobility of the 2D hole gas (2DHG) at the GaN/AlN interface, and evaluates the potential of strain-engineering approaches to alter the bandstructure in a favorable way.  First, the valence band structure including confinement and multi-band mixing effects are computed.  Then the spectra of both acoustic and optical phonons in the heterostructure are determined and the mobility limitation due to these mechanisms is derived.  Finally, the effects of strain on the band structure and mobility are presented.

The multi-band k.p (MBKP) approach, based on Burt Exact Envelope Function Theory \cite{Foreman1993,Burt1994}, describes the electronic states of heterostructures wherein multiple subbands may be intermixed by non-uniform potentials and material interfaces. MBKP has been extended to wurtzite heterostructures by various authors \cite{Mireles1999,Chuang1996} and is available in certain commercial packages \cite{BirnerThesis}.  The PyNitride software package \cite{PyNitride} employed here self-consistently \cite{Tan1990} solves the 6x6 MBKP equation of the wurtzite valence band (see Supplement) simultaneously with the Poisson equation, accounting for the large fixed interface polarization charge.
Figure \ref{fig:2DHG}(a) depicts a highly confined 2D hole gas at the GaN/AlN interface.  The hole gas represents contributions mainly from the first quantized subband of both the heavy hole (HH) and light hole (LH) bands (including spin, this is four subbands). The transverse dispersion is shown in Fig \ref{fig:2DHG}(b).  By density, as clear from Fig \ref{fig:2DHG}(c), the HH band dominates, though as shown in Fig \ref{fig:2DHG}(d), the LH band has lighter in-plane mass near the Fermi energy, so it contributes significantly to transport.  To evaluate transport, we proceed to describe the phonons.

The Dielectric Continuum model\cite{StroscioDutta} describes polar optical phonons (POPs) in arbitrary heterostructures.  As POP scattering is the main limitation on electron mobility in quality GaN, numerous authors have invested significant theoretical effort into the elaboration of POP spectra in various wurtzite heterostructures \cite{Shi2003, Medeiros2005, Liao2010, Komirenko1999a, Zhu2012}. A structure as simple as this, in fact, can be solved analytically.  For a uniaxial crystal, the effect of polar optical phonons can be incorporated into two frequency ($\omega$) dependent, directional dielectric functions, $\varepsilon_\perp$ and $\varepsilon_\parallel$:
\begin{equation}
  \varepsilon_\perp=\varepsilon^\infty\frac{\omega_{LO\perp}^2-                 \omega^2}{\omega_{TO\perp}^2-\omega^2},
  \quad\mathrm{and}\quad
  \varepsilon_\parallel=\varepsilon^\infty\frac{\omega_{LO\parallel}^2-         \omega^2}{\omega_{TO\parallel}^2-\omega^2},
  \label{eq:diels}
\end{equation}
where those longitudinal ($\omega_{LO}$) and transverse ($\omega_{TO}$) POP frequencies and high-frequency dielectric constant $\epsilon^\infty$ can be determined experimentally for the materials in play (see tabulation in Komirenko \cite{Komirenko1999a}). For a mode with in-plane wavevector $q$, the POP problem reduces to solving a frequency-dependent Poisson eigenvalue equation:
\begin{equation}
  \partial_z\epsilon_\parallel\partial_z\phi=q^2\epsilon_\perp\phi
  \label{eq:poissoneig}
\end{equation}
for the potential $\phi$. At every characteristic frequency appearing in Eq \ref{eq:diels}, a dielectric constant changes sign, which changes the character of the involved modes (locally decaying vs oscillating).  Altogether, there are three classes of modes depending on the energy range: (1) oscillating in GaN, decaying in AlN, (2) oscillating in AlN, decaying in GaN, and (3) decaying bidirectionally from a GaN/AlN interface.  For each class, both transverse and longitudinal polarizations are possible.  The spectrum and example modes are depicted in Fig \ref{fig:pop} and the solution is elaborated in the Supplementary Materials.

The Elastic Continuum model \cite{StroscioDutta} aptly describes acoustic phonons in arbitrary heterostructures.  Given the centrality of \textit{optical} phonons in electron-based devices, the literature on acoustic phonons in wurtzite heterostructures \cite{Pokatilov2003,Pokatilov2004,Balandin2004,Balandin2007} is significantly less comprehensive. The basic approach is to link a continuum Newton's law with a material stress-strain relation:
\begin{equation}
  \rho \frac{\partial^2 u_i}{\partial t^2}=\frac{\partial T_{ij}}{\partial r_j}, \quad T_{ij}=c_{ijkl}\epsilon_{kl},
\end{equation}
where $u_i$ is the local displacement vector, $\rho$ is the density, $T_{ij}$ is the stress tensor, $c_{ijkl}$ is the stiffness tensor, and $\epsilon_{ijkl}$ is the strain tensor $\epsilon_{ij}=\frac{1}{2}\left( \partial_{r_j} u_i + \partial_{r_i} u_j\right)$.  Due to symmetry constraints on the stiffness tensor, the in-plane shear component is uncoupled (``Y'' modes) while longitudinal and out-of-plane components constitute two coupled differential equations (``XZ'' modes). These are solved by the Finite Element Method.
For both types of modes, there are two characters possible depending on the energy range.  At energies above $\hbar v_{t}q$, where $v_t$ is the relevant AlN sound velocity, modes are able to propagate in the AlN, and are thus of a continuum sort.  Below this energy, any modes which exist must decay into the AlN, so are GaN-confined modes.  The spectrum and example modes are revealed in Fig \ref{fig:acoustic}, and the solution methodology is elaborated in the Supplementary Materials.

\begin{figure}[t]
  \centering
  \includegraphics[width=\columnwidth]{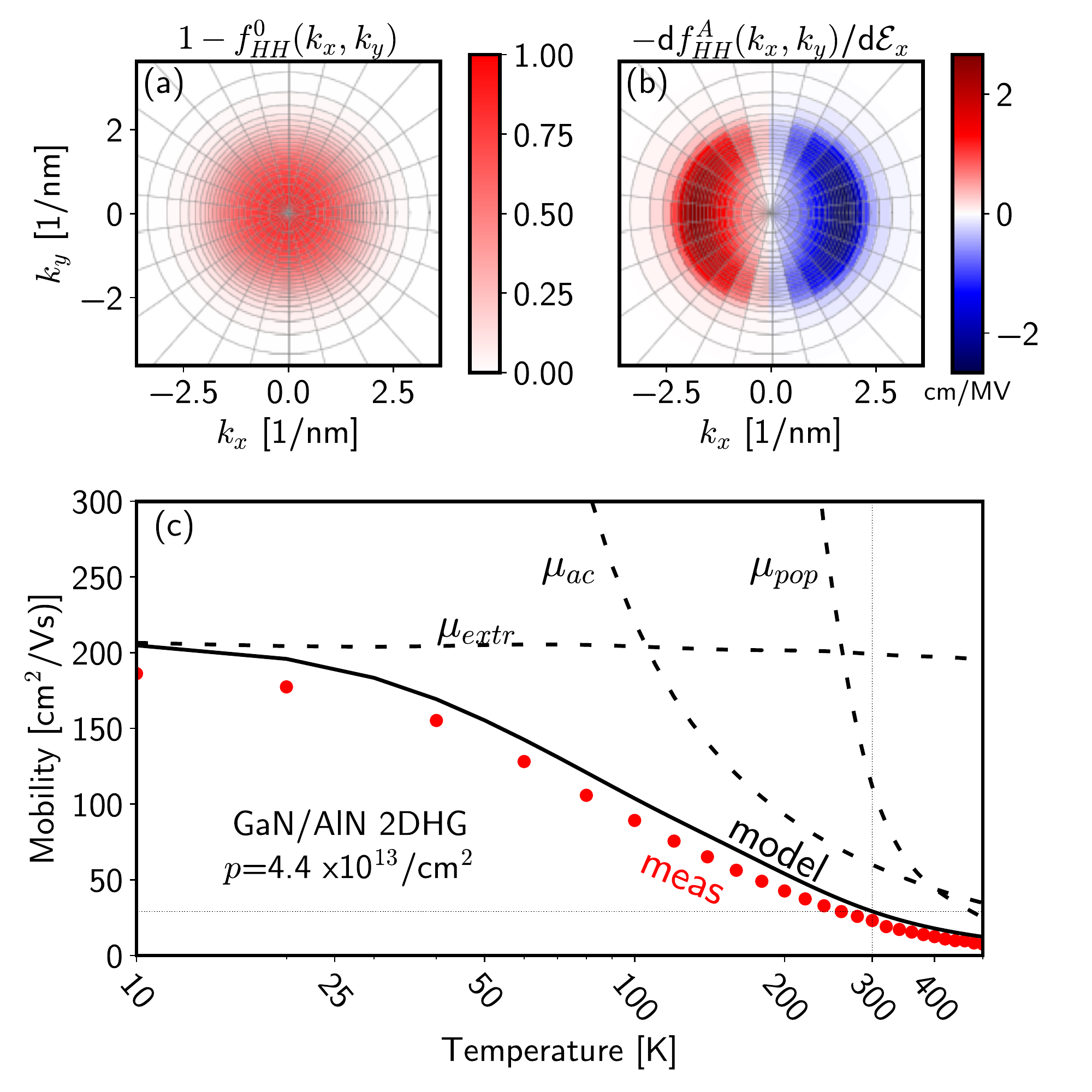}
  \caption{(a) The hole occupation of the HH band at equilibrium and (b) the antisymmetric change in occuption per applied electric field in the $+x$ direction, obtained by direct solution of the linearized Boltzmann Transport Equation. (The hole population is enhanced at negative $k_x$, which, given the negative group velocity for holes, implies a rightward current.) (c) Model mobility versus Hall measurements reported by Chaudhuri et al. \cite{Chaudhuri2018} Dashed durves are also shown for the various scattering mechanisms alone (polar optical, acoustic, and extrinsic). Note: the model is obtained by a full solution incorporating all mechanisms simultaneously, not by a Matthiessen approximation of the component limitations. }
  \label{fig:model}
\end{figure}
We now turn to the interactions of the phonons and carriers. First, both types of phonon modes are normalized\cite{StroscioDutta} by the quantization condition
  $\int d^3 \vec{r}\ \rho(\vec r)\abs{\vec u(r)}^2= \frac{\hbar}{2\omega}$.
  Once the oscillations are normalized, the Hamiltonian term for each mode can be generated. For optical phonons, this is just $H_{pop}=-e\phi(r)$.  For acoustic phonons, the deformation potential $H_{adp}=D(\epsilon)$ is a six-by-six matrix function of position: at each position it recruits the valence deformation matrix of the local material and also the local strain from a given acoustic mode.  There is also a (coherently-combined) piezo term $H_{pz}=-e\phi(r)$ where $\phi$ is found\cite{Pokatilov2004} by solving the Poisson equation given the piezoelectric charge induced by the mode (see Supplement).  Finally, to account for the low-temperature mobility, which is around 200cm$^2$/Vs in these structures to date, some extrinsic limitation must be included.  The exact cause (\textit{e.g.} interface roughness, dislocation, impurity, etc) is irrelevant to this study, since all these elastic mechanisms are temperature-independent and have similar dependence on effective masses.  So the exact cause is not deduced here, but rather a generic scatterer with constant scalar matrix element is applied to set the low-temperature mobility to 200cm$^2$/Vs.  This matrix element is the only ``tuning parameter.''

To calculate the perturbed carrier distribution, we employ the linear Boltzmann Transport Equation (LBTE):
\begin{multline}
  \frac{q\vec{\mathcal{E}}}{\hbar}\cdot\nabla_kf^0_m(k)=
  \sum_{k'm'} R^{k'm'}_{km}f_m^A(k)-R^{km}_{k'm'}f_{m'}^A(k')
  \label{eq:LBTE}
\end{multline}
where $\vec{\mathcal{E}}$ is the in-plane electric field, $f^0$ is the equilbrium occupation function, $f^A$ is a small change in occupation to be solved for, and the transition rate $R$ is given by 
\begin{multline}
R^{k'm'}_{km}=\frac{2\pi}{\hbar}\sum_{\alpha l}\abs{\bra{\psi_{k'm'}}H^l(q)\ket{\psi_{km}}}^2\\
  \times\left(N+\frac{1}{2}\left( 1-\alpha \right)+\alpha f^0_{m'}(k')\right)\\
  \times\delta(E_{k'm'}-E_{km}-\alpha \varepsilon_l(q)),
\end{multline}
where $\alpha=\pm 1$ represents absorption/emission respectively, $\psi_{km}$ is the state with in-plane wavevector $\vec k$, of subband $m$, $H^l(q)$ is the perturbation due to mode $l$ with wavevector $\vec q=\vec k'-\vec k$, and $E$/$\varepsilon$ are the electronic/phonon energies.  Once this is discretized on a k-space mesh, the LBTE is solved as a linear matrix equation for the change in occupation $\partial f^A/\partial\mathcal{E}$ under applied field, from which the mobility is extracted.  The results, Fig \ref{fig:model}, compare agreeably to experiment over a wide temperature range, and verify Ponc\'e's prediction\cite{Ponce2019} that acoustic phonons dominate scattering at room temperature.

Now we discuss what can be done to improve the mobility.  In a recent first-principles study of the mobility of bulk p-type GaN, Ponc\'e et al \cite{Ponce2019} suggested the application of significant tensile in-plane strain (or compressive c-axis strain) to raise the split-off band (SO) above the heavy-hole and light-hole bands.  The lighter mass of the split-off band would allow for a drastically increased hole mobility.  Such a proposal, while potentially revolutionary for bulk p-GaN, is difficult to apply to the particular heterostructure discussed here-in, since the GaN-pseudomorphic-to-AlN is already \textit{compressively} strained by 2.4\% as-grown, and that large strain would have to be overcome first before applying further tensile strain.  Moreover, applying tensile strain dramatically changes the interface between GaN and AlN, significantly reducing the valence band offset (VBO) and thus the confining potential.  [The role of strain in the asymmetry of the GaN/AlN and AlN/GaN VBO is well-known \cite{VM2003}.]

\begin{figure}[t]
  \centering
  \includegraphics[width=\columnwidth]{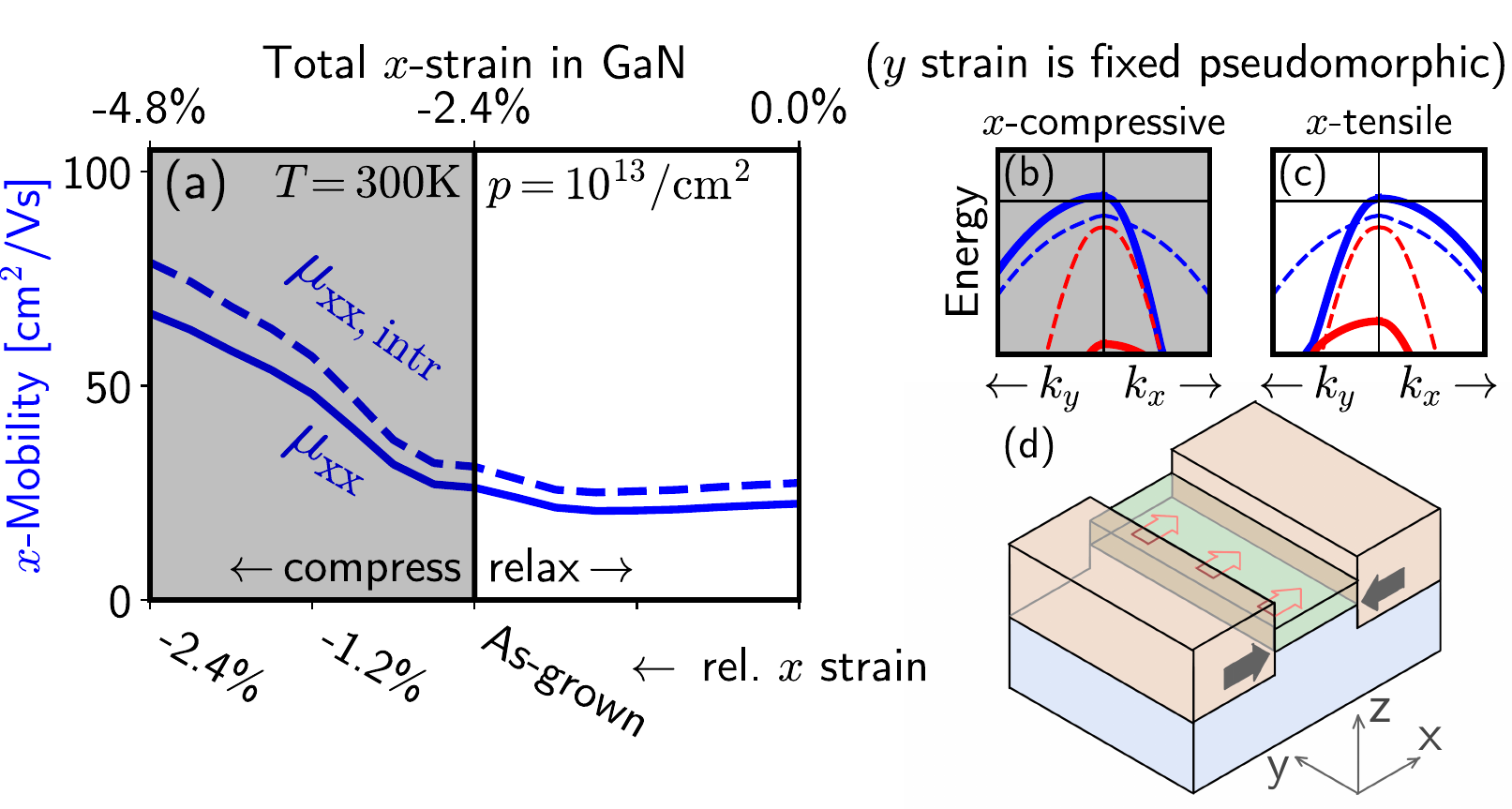}
  \caption{(a) $x$-directed mobility due to all mechanisms (solid blue) and phonons only (dashed blue) versus strain along the current-flow direction ($x$). The $y$-direction is left pseudomorphic (-2.4\% $y$-strain in GaN).  Bottom axis: $x$-strain applied to structure. Top axis: corresponding total $x$-strain in GaN.  For compressive $x$-strain, the $x$-directed mobility is enhanced. The dispersion along is shown under (b) compression and (c) tension as solid curves (dashed as-grown bands provided for comparison). The $x$-directed mobility enhancement corresponds to a reduced effective mass along $k_x$ of the topmost band in (b). (d) Such compression could be applied for example by regrown stressors on the source and drain.}
  \label{fig:xstrain}
  \centering
  \includegraphics[width=\columnwidth]{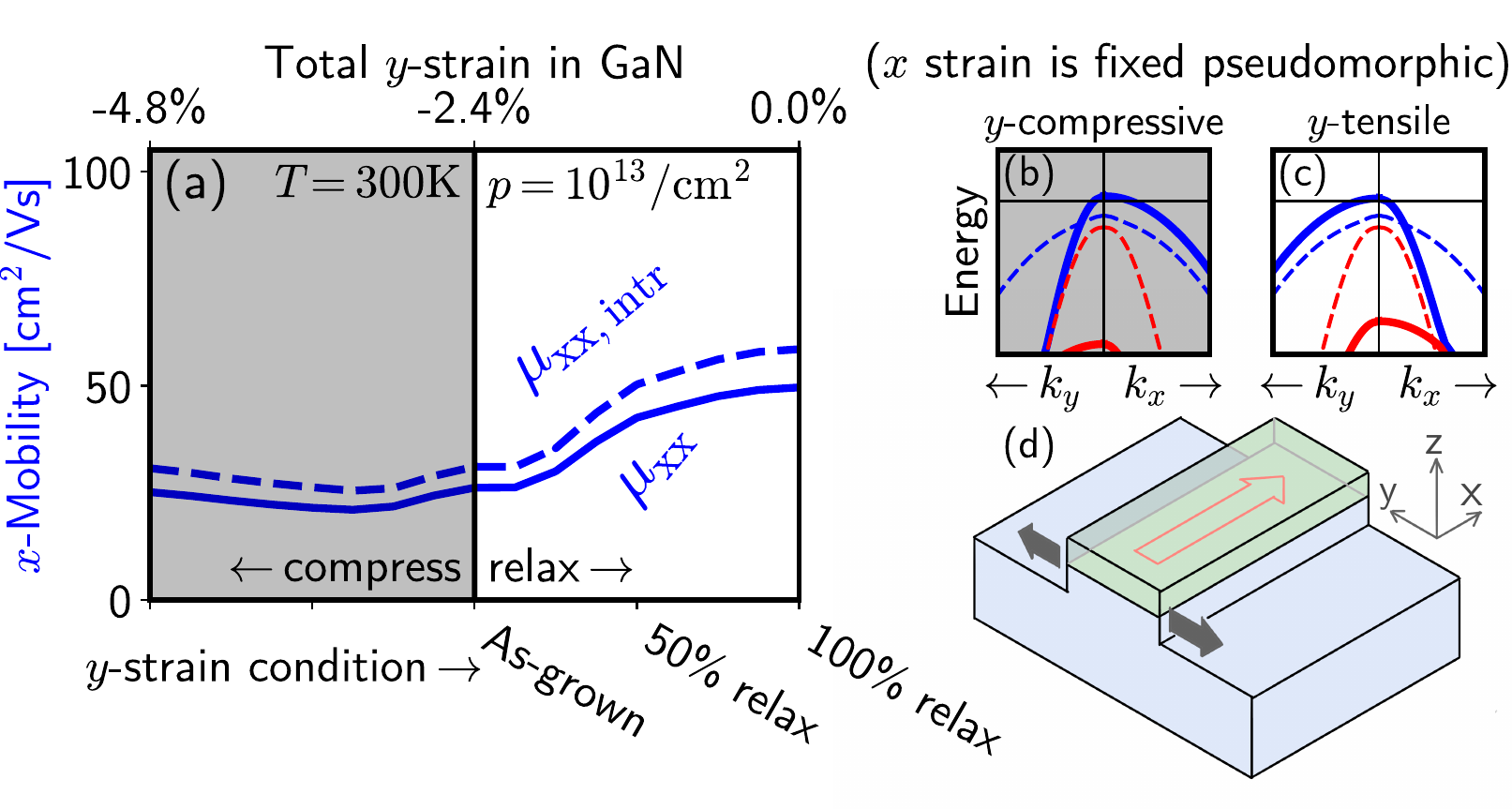}
  \caption{Similar display as Fig \ref{fig:xstrain} but for strain applied \textit{perpendicular} ($y$) to current flow ($x$).  The $x$-direction is left pseudomorphic (-2.4\% $x$-strain in GaN).  (a) Tensile $y$-strain provides a mobility enhancement, corresponding to (c) a lighter effective mass along $x$.  The bottom axis of (a) describes the tensile condition in terms of a unidirectional relaxation of the GaN strain perpendicular to current flow, as indicated by the fin release geometry of (d).}
  \label{fig:ystrain}
\end{figure}

This raises the question of whether reasonable strain can improve the mobility of holes in this heterostructure, without access to the deep SO band.  
Based on the theory of Suzuki and Oenoyama \cite{Suzuki1996} in the context of lasers, two converse techniques may be suggested employing strain along \textit{only one in-plane axis}.
Dasgupta et al \cite{DasguptaPatent2017} considered the application of \textit{compressive} strain \textit{along} the direction of current flow. Conversely, Gupta et al \cite{Gupta2018} tested the application of \textit{tensile} strain \textit{perpendicular} to current flow.
Figures \ref{fig:xstrain} and \ref{fig:ystrain} consider what effects these two proposals have, not on bulk p-GaN as with previous authors, but rather on the heterojunction band structure.  To take full advantage of the changing band structure near the band edge, the charge density is lowered [by thinning the well layer to 8nm] and kept constant at $1\times 10^{13}/\mathrm{cm}^2$ by variable applied bias even as the strain is changed.  These adjustments to thickness and charge match closely the actual 2DHG environment in p-channel FETs \cite{BaderEDL2018}.  In parts (b) and (c) of these figures, uniaxial in-plane strain is seen to split the topmost two-bands.  Depending on the sign and orientation of the strain, the topmost band may become light or heavy in the current flow direction ($x$).

In the case of tensile $x$-strain, Fig \ref{fig:xstrain}(c), or compressive $y$-strain, Fig \ref{fig:ystrain}(b), the topmost band is heavy along $x$, so, even though the available scattering DOS is reduced by the band-splitting, the net mobility remains low.  However, in the case of compressive $x$-strain, Fig \ref{fig:xstrain}(b), or tensile $y$ strain, Fig \ref{fig:ystrain}(c), the topmost band is light along $x$ \textit{and} interband scattering is diminished.  As such, these strain conditions result in an enhance mobility.  The mobility improvement due to $x$-compressive strain is more pronounced than $y$-tensile strain because of the aforementioned effect which tensile strain has on the VBO. However, these two enhancement mechanisms are not mutually exclusive: a fin relaxation and a compressive regrowth would complement each other via the Poisson effect to reduce the total amount of stress which each mechanism would have to apply.

Of course, any enhancements will still be limited by whatever extrinsic mechanisms are at play, so Figs \ref{fig:xstrain} \& \ref{fig:ystrain} include two mobility curves: (1) a (solid blue) prediction which assumes the extrinsic scattering element is similar to the present data and (2) a (dashed blue) intrinsic calculation using only phonon mechanisms, corresponding to space for further improvements in growth.

In conclusion, this work has combined solutions for both the optical and acoustic phonon spectra in a GaN/AlN heterostructure with a multiband description of the hole gas to model the hole mobility.  The model qualitatively matches experimental observations and can be used to estimate what sort of intrinsic mobility enhancements are possible by strain engineering, suggesting that a unixial compression is the best approach for optimizing the GaN/AlN hole gas.  A high-mobility 2DHG could join the celebrated GaN 2DEG to realize a future of energy-efficient complementary GaN-based circuits.

\section*{Supplementary Material}
See the Supplementary Material for mathematical and numerical simulation details.

%
%

%

\begin{acknowledgments}
  Work supported by Intel Corp, by AFOSR Grant FA9550-17-1-0048, and by NSF Grants 1710298/1534303. Authors appreciate discussions with Samuel Ponc\'e, Feliciano Giustino, Tom\'as Palacios, Nadim Chowdhury, Paul Fischer and Sansaptak Dasgupta.
\end{acknowledgments}

\bibliography{Pubs-2019Mobility}

\begin{thebibliography}{38}%
\makeatletter
\providecommand \@ifxundefined [1]{%
 \@ifx{#1\undefined}
}%
\providecommand \@ifnum [1]{%
 \ifnum #1\expandafter \@firstoftwo
 \else \expandafter \@secondoftwo
 \fi
}%
\providecommand \@ifx [1]{%
 \ifx #1\expandafter \@firstoftwo
 \else \expandafter \@secondoftwo
 \fi
}%
\providecommand \natexlab [1]{#1}%
\providecommand \enquote  [1]{``#1''}%
\providecommand \bibnamefont  [1]{#1}%
\providecommand \bibfnamefont [1]{#1}%
\providecommand \citenamefont [1]{#1}%
\providecommand \href@noop [0]{\@secondoftwo}%
\providecommand \href [0]{\begingroup \@sanitize@url \@href}%
\providecommand \@href[1]{\@@startlink{#1}\@@href}%
\providecommand \@@href[1]{\endgroup#1\@@endlink}%
\providecommand \@sanitize@url [0]{\catcode `\\12\catcode `\$12\catcode
  `\&12\catcode `\#12\catcode `\^12\catcode `\_12\catcode `\%12\relax}%
\providecommand \@@startlink[1]{}%
\providecommand \@@endlink[0]{}%
\providecommand \url  [0]{\begingroup\@sanitize@url \@url }%
\providecommand \@url [1]{\endgroup\@href {#1}{\urlprefix }}%
\providecommand \urlprefix  [0]{URL }%
\providecommand \Eprint [0]{\href }%
\providecommand \doibase [0]{http://dx.doi.org/}%
\providecommand \selectlanguage [0]{\@gobble}%
\providecommand \bibinfo  [0]{\@secondoftwo}%
\providecommand \bibfield  [0]{\@secondoftwo}%
\providecommand \translation [1]{[#1]}%
\providecommand \BibitemOpen [0]{}%
\providecommand \bibitemStop [0]{}%
\providecommand \bibitemNoStop [0]{.\EOS\space}%
\providecommand \EOS [0]{\spacefactor3000\relax}%
\providecommand \BibitemShut  [1]{\csname bibitem#1\endcsname}%
\let\auto@bib@innerbib\@empty
\bibitem [{\citenamefont {Nakamura}\ \emph {et~al.}(1992)\citenamefont
  {Nakamura}, \citenamefont {Mukai}, \citenamefont {Senoh},\ and\ \citenamefont
  {Iwasa}}]{Shuji1992}%
  \BibitemOpen
  \bibfield  {author} {\bibinfo {author} {\bibfnamefont {S.}~\bibnamefont
  {Nakamura}}, \bibinfo {author} {\bibfnamefont {T.}~\bibnamefont {Mukai}},
  \bibinfo {author} {\bibfnamefont {M.}~\bibnamefont {Senoh}}, \ and\ \bibinfo
  {author} {\bibfnamefont {N.}~\bibnamefont {Iwasa}},\ }\bibfield  {title}
  {\enquote {\bibinfo {title} {{Thermal Annealing Effects on P-Type Mg-Doped
  GaN Films}},}\ }\href {\doibase 10.1143/JJAP.31.L139} {\bibfield  {journal}
  {\bibinfo  {journal} {Japanese J. App. Phys.}\ }\textbf {\bibinfo {volume}
  {31}},\ \bibinfo {pages} {L139} (\bibinfo {year} {1992})}\BibitemShut
  {NoStop}%
\bibitem [{\citenamefont {Huang}(2017)}]{Huang2017}%
  \BibitemOpen
  \bibfield  {author} {\bibinfo {author} {\bibfnamefont {A.~Q.}\ \bibnamefont
  {Huang}},\ }\bibfield  {title} {\enquote {\bibinfo {title} {{Wide bandgap
  (WBG) power devices and their impacts on power delivery systems}},}\ }\href
  {\doibase 10.1109/IEDM.2016.7838457} {\bibfield  {journal} {\bibinfo
  {journal} {Technical Digest - Intl. Electron Devices Meeting, IEDM}\ ,\
  \bibinfo {pages} {20.1.1.1.4}} (\bibinfo {year} {2017})}\BibitemShut
  {NoStop}%
\bibitem [{\citenamefont {Yuk}, \citenamefont {Branner},\ and\ \citenamefont
  {Cui}(2017)}]{Yuk2017}%
  \BibitemOpen
  \bibfield  {author} {\bibinfo {author} {\bibfnamefont {K.}~\bibnamefont
  {Yuk}}, \bibinfo {author} {\bibfnamefont {G.~R.}\ \bibnamefont {Branner}}, \
  and\ \bibinfo {author} {\bibfnamefont {C.}~\bibnamefont {Cui}},\ }\bibfield
  {title} {\enquote {\bibinfo {title} {{Future directions for GaN in 5G and
  satellite communications}},}\ }\href {\doibase 10.1109/MWSCAS.2017.8053045}
  {\bibfield  {journal} {\bibinfo  {journal} {Midwest Symposium on Circuits and
  Systems}\ }\textbf {\bibinfo {volume} {2017}},\ \bibinfo {pages} {803}
  (\bibinfo {year} {2017})}\BibitemShut {NoStop}%
\bibitem [{\citenamefont {Song}, \citenamefont {Ha},\ and\ \citenamefont
  {Seong}(2010)}]{Song2010}%
  \BibitemOpen
  \bibfield  {author} {\bibinfo {author} {\bibfnamefont {J.~O.}\ \bibnamefont
  {Song}}, \bibinfo {author} {\bibfnamefont {J.~S.}\ \bibnamefont {Ha}}, \ and\
  \bibinfo {author} {\bibfnamefont {T.~Y.}\ \bibnamefont {Seong}},\ }\bibfield
  {title} {\enquote {\bibinfo {title} {{Ohmic-contact technology for GaN-based
  light-emitting diodes: Role of p-type contact}},}\ }\href {\doibase
  10.1109/TED.2009.2034506} {\bibfield  {journal} {\bibinfo  {journal} {IEEE
  Transactions on Electron Devices}\ }\textbf {\bibinfo {volume} {57}},\
  \bibinfo {pages} {42} (\bibinfo {year} {2010})}\BibitemShut {NoStop}%
\bibitem [{\citenamefont {Kozodoy}\ \emph {et~al.}(2000)\citenamefont
  {Kozodoy}, \citenamefont {Xing}, \citenamefont {DenBaars}, \citenamefont
  {Mishra}, \citenamefont {Saxler}, \citenamefont {Perrin}, \citenamefont
  {Elhamri},\ and\ \citenamefont {Mitchel}}]{Kozodoy2000}%
  \BibitemOpen
  \bibfield  {author} {\bibinfo {author} {\bibfnamefont {P.}~\bibnamefont
  {Kozodoy}}, \bibinfo {author} {\bibfnamefont {H.}~\bibnamefont {Xing}},
  \bibinfo {author} {\bibfnamefont {S.~P.}\ \bibnamefont {DenBaars}}, \bibinfo
  {author} {\bibfnamefont {U.~K.}\ \bibnamefont {Mishra}}, \bibinfo {author}
  {\bibfnamefont {A.}~\bibnamefont {Saxler}}, \bibinfo {author} {\bibfnamefont
  {R.}~\bibnamefont {Perrin}}, \bibinfo {author} {\bibfnamefont
  {S.}~\bibnamefont {Elhamri}}, \ and\ \bibinfo {author} {\bibfnamefont
  {W.~C.}\ \bibnamefont {Mitchel}},\ }\bibfield  {title} {\enquote {\bibinfo
  {title} {{Heavy doping effects in Mg-doped GaN}},}\ }\href {\doibase
  10.1063/1.372098} {\bibfield  {journal} {\bibinfo  {journal} {J. App. Phys.}\
  }\textbf {\bibinfo {volume} {87}},\ \bibinfo {pages} {1832} (\bibinfo {year}
  {2000})}\BibitemShut {NoStop}%
\bibitem [{\citenamefont {Chu}\ \emph {et~al.}(2016)\citenamefont {Chu},
  \citenamefont {Cao}, \citenamefont {Chen}, \citenamefont {Li},\ and\
  \citenamefont {Zehnder}}]{Chu2016}%
  \BibitemOpen
  \bibfield  {author} {\bibinfo {author} {\bibfnamefont {R.}~\bibnamefont
  {Chu}}, \bibinfo {author} {\bibfnamefont {Y.}~\bibnamefont {Cao}}, \bibinfo
  {author} {\bibfnamefont {M.}~\bibnamefont {Chen}}, \bibinfo {author}
  {\bibfnamefont {R.}~\bibnamefont {Li}}, \ and\ \bibinfo {author}
  {\bibfnamefont {D.}~\bibnamefont {Zehnder}},\ }\bibfield  {title} {\enquote
  {\bibinfo {title} {{An Experimental Demonstration of GaN CMOS Technology}},}\
  }\href {\doibase 10.1109/LED.2016.2515103} {\bibfield  {journal} {\bibinfo
  {journal} {IEEE Elec. Dev. Lett.}\ }\textbf {\bibinfo {volume} {37}},\
  \bibinfo {pages} {269} (\bibinfo {year} {2016})}\BibitemShut {NoStop}%
\bibitem [{\citenamefont {Chaudhuri}\ \emph {et~al.}()\citenamefont
  {Chaudhuri}, \citenamefont {Bader}, \citenamefont {Chen}, \citenamefont
  {Muller}, \citenamefont {Xing},\ and\ \citenamefont {Jena}}]{Chaudhuri2018}%
  \BibitemOpen
  \bibfield  {author} {\bibinfo {author} {\bibfnamefont {R.}~\bibnamefont
  {Chaudhuri}}, \bibinfo {author} {\bibfnamefont {S.~J.}\ \bibnamefont
  {Bader}}, \bibinfo {author} {\bibfnamefont {Z.}~\bibnamefont {Chen}},
  \bibinfo {author} {\bibfnamefont {D.~A.}\ \bibnamefont {Muller}}, \bibinfo
  {author} {\bibfnamefont {H.~G.}\ \bibnamefont {Xing}}, \ and\ \bibinfo
  {author} {\bibfnamefont {D.}~\bibnamefont {Jena}},\ }\bibfield  {title}
  {\enquote {\bibinfo {title} {{A polarization-induced 2D hole gas in undoped
  gallium nitride quantum wells}},}\ }\href@noop {} {\bibfield  {journal}
  {\bibinfo  {journal} {Unpublished (in review)}\ }}\Eprint
  {http://arxiv.org/abs/1807.08836} {arXiv:1807.08836} \BibitemShut {NoStop}%
\bibitem [{\citenamefont {Bader}\ \emph {et~al.}(2018)\citenamefont {Bader},
  \citenamefont {Chaudhuri}, \citenamefont {Nomoto}, \citenamefont {Hickman},
  \citenamefont {Chen}, \citenamefont {Then}, \citenamefont {Muller},
  \citenamefont {Xing},\ and\ \citenamefont {Jena}}]{BaderEDL2018}%
  \BibitemOpen
  \bibfield  {author} {\bibinfo {author} {\bibfnamefont {S.~J.}\ \bibnamefont
  {Bader}}, \bibinfo {author} {\bibfnamefont {R.}~\bibnamefont {Chaudhuri}},
  \bibinfo {author} {\bibfnamefont {K.}~\bibnamefont {Nomoto}}, \bibinfo
  {author} {\bibfnamefont {A.}~\bibnamefont {Hickman}}, \bibinfo {author}
  {\bibfnamefont {Z.}~\bibnamefont {Chen}}, \bibinfo {author} {\bibfnamefont
  {H.~W.}\ \bibnamefont {Then}}, \bibinfo {author} {\bibfnamefont {D.~A.}\
  \bibnamefont {Muller}}, \bibinfo {author} {\bibfnamefont {H.~G.}\
  \bibnamefont {Xing}}, \ and\ \bibinfo {author} {\bibfnamefont
  {D.}~\bibnamefont {Jena}},\ }\bibfield  {title} {\enquote {\bibinfo {title}
  {{Gate-recessed E-mode p-channel HFET with high on-current based on GaN/AlN
  2D hole gas}},}\ }\href {\doibase 10.1109/LED.2018.2874190} {\bibfield
  {journal} {\bibinfo  {journal} {IEEE Elec. Dev. Lett.}\ }\textbf {\bibinfo
  {volume} {39}},\ \bibinfo {pages} {1848} (\bibinfo {year}
  {2018})}\BibitemShut {NoStop}%
\bibitem [{\citenamefont {Shatalov}\ \emph {et~al.}(2002)\citenamefont
  {Shatalov}, \citenamefont {Simin}, \citenamefont {Zhang}, \citenamefont
  {Adivarahan}, \citenamefont {Koudymov}, \citenamefont {Pachipulusu},\ and\
  \citenamefont {Khan}}]{Shatalov2002}%
  \BibitemOpen
  \bibfield  {author} {\bibinfo {author} {\bibfnamefont {M.}~\bibnamefont
  {Shatalov}}, \bibinfo {author} {\bibfnamefont {G.}~\bibnamefont {Simin}},
  \bibinfo {author} {\bibfnamefont {J.}~\bibnamefont {Zhang}}, \bibinfo
  {author} {\bibfnamefont {V.}~\bibnamefont {Adivarahan}}, \bibinfo {author}
  {\bibfnamefont {A.}~\bibnamefont {Koudymov}}, \bibinfo {author}
  {\bibfnamefont {R.}~\bibnamefont {Pachipulusu}}, \ and\ \bibinfo {author}
  {\bibfnamefont {M.~A.}\ \bibnamefont {Khan}},\ }\bibfield  {title} {\enquote
  {\bibinfo {title} {{GaN/AlGaN p-channel inverted heterostructure JFET}},}\
  }\href {\doibase 10.1109/LED.2002.801295} {\bibfield  {journal} {\bibinfo
  {journal} {IEEE Elec. Dev. Lett.}\ }\textbf {\bibinfo {volume} {23}},\
  \bibinfo {pages} {452} (\bibinfo {year} {2002})}\BibitemShut {NoStop}%
\bibitem [{\citenamefont {Zimmermann}\ \emph {et~al.}(2004)\citenamefont
  {Zimmermann}, \citenamefont {Neuburger}, \citenamefont {Kunze}, \citenamefont
  {Daumiller}, \citenamefont {Denisenko}, \citenamefont {Dadgar}, \citenamefont
  {Krost},\ and\ \citenamefont {Kohn}}]{Zimmermann2004}%
  \BibitemOpen
  \bibfield  {author} {\bibinfo {author} {\bibfnamefont {T.}~\bibnamefont
  {Zimmermann}}, \bibinfo {author} {\bibfnamefont {M.}~\bibnamefont
  {Neuburger}}, \bibinfo {author} {\bibfnamefont {M.}~\bibnamefont {Kunze}},
  \bibinfo {author} {\bibfnamefont {I.}~\bibnamefont {Daumiller}}, \bibinfo
  {author} {\bibfnamefont {A.}~\bibnamefont {Denisenko}}, \bibinfo {author}
  {\bibfnamefont {A.}~\bibnamefont {Dadgar}}, \bibinfo {author} {\bibfnamefont
  {A.}~\bibnamefont {Krost}}, \ and\ \bibinfo {author} {\bibfnamefont
  {E.}~\bibnamefont {Kohn}},\ }\bibfield  {title} {\enquote {\bibinfo {title}
  {{P-channel InGaN-HFET structure based on polarization doping}},}\ }\href
  {\doibase 10.1109/DRC.2003.1226852} {\bibfield  {journal} {\bibinfo
  {journal} {IEEE Elec. Dev. Lett.}\ }\textbf {\bibinfo {volume} {25}},\
  \bibinfo {pages} {450} (\bibinfo {year} {2004})}\BibitemShut {NoStop}%
\bibitem [{\citenamefont {Li}\ \emph {et~al.}(2013)\citenamefont {Li},
  \citenamefont {Wang}, \citenamefont {Song}, \citenamefont {Verma},
  \citenamefont {Cao}, \citenamefont {Ganguly}, \citenamefont {Verma},
  \citenamefont {Guo}, \citenamefont {Xing},\ and\ \citenamefont
  {Jena}}]{Li2013}%
  \BibitemOpen
  \bibfield  {author} {\bibinfo {author} {\bibfnamefont {G.}~\bibnamefont
  {Li}}, \bibinfo {author} {\bibfnamefont {R.}~\bibnamefont {Wang}}, \bibinfo
  {author} {\bibfnamefont {B.}~\bibnamefont {Song}}, \bibinfo {author}
  {\bibfnamefont {J.}~\bibnamefont {Verma}}, \bibinfo {author} {\bibfnamefont
  {Y.}~\bibnamefont {Cao}}, \bibinfo {author} {\bibfnamefont {S.}~\bibnamefont
  {Ganguly}}, \bibinfo {author} {\bibfnamefont {A.}~\bibnamefont {Verma}},
  \bibinfo {author} {\bibfnamefont {J.}~\bibnamefont {Guo}}, \bibinfo {author}
  {\bibfnamefont {H.~G.}\ \bibnamefont {Xing}}, \ and\ \bibinfo {author}
  {\bibfnamefont {D.}~\bibnamefont {Jena}},\ }\bibfield  {title} {\enquote
  {\bibinfo {title} {{Polarization-induced GaN-on-insulator E/D Mode p-channel
  heterostructure FETs}},}\ }\href {\doibase 10.1109/LED.2013.2264311}
  {\bibfield  {journal} {\bibinfo  {journal} {IEEE Elec. Dev. Lett.}\ }\textbf
  {\bibinfo {volume} {34}},\ \bibinfo {pages} {852} (\bibinfo {year}
  {2013})}\BibitemShut {NoStop}%
\bibitem [{\citenamefont {Hahn}\ \emph {et~al.}(2013)\citenamefont {Hahn},
  \citenamefont {Reuters}, \citenamefont {Pooth}, \citenamefont {Hollander},
  \citenamefont {Heuken}, \citenamefont {Kalisch},\ and\ \citenamefont
  {Vescan}}]{Hahn2013}%
  \BibitemOpen
  \bibfield  {author} {\bibinfo {author} {\bibfnamefont {H.}~\bibnamefont
  {Hahn}}, \bibinfo {author} {\bibfnamefont {B.}~\bibnamefont {Reuters}},
  \bibinfo {author} {\bibfnamefont {A.}~\bibnamefont {Pooth}}, \bibinfo
  {author} {\bibfnamefont {B.}~\bibnamefont {Hollander}}, \bibinfo {author}
  {\bibfnamefont {M.}~\bibnamefont {Heuken}}, \bibinfo {author} {\bibfnamefont
  {H.}~\bibnamefont {Kalisch}}, \ and\ \bibinfo {author} {\bibfnamefont
  {A.}~\bibnamefont {Vescan}},\ }\bibfield  {title} {\enquote {\bibinfo {title}
  {{P-channel enhancement and depletion mode GaN-based HFETs with quaternary
  backbarriers}},}\ }\href {\doibase 10.1109/TED.2013.2272330} {\bibfield
  {journal} {\bibinfo  {journal} {IEEE Transactions on Electron Devices}\
  }\textbf {\bibinfo {volume} {60}},\ \bibinfo {pages} {3005} (\bibinfo {year}
  {2013})}\BibitemShut {NoStop}%
\bibitem [{\citenamefont {Zhang}\ \emph {et~al.}(2016)\citenamefont {Zhang},
  \citenamefont {Sumiya}, \citenamefont {Liao}, \citenamefont {Koide},\ and\
  \citenamefont {Sang}}]{Zhang2016}%
  \BibitemOpen
  \bibfield  {author} {\bibinfo {author} {\bibfnamefont {K.}~\bibnamefont
  {Zhang}}, \bibinfo {author} {\bibfnamefont {M.}~\bibnamefont {Sumiya}},
  \bibinfo {author} {\bibfnamefont {M.}~\bibnamefont {Liao}}, \bibinfo {author}
  {\bibfnamefont {Y.}~\bibnamefont {Koide}}, \ and\ \bibinfo {author}
  {\bibfnamefont {L.}~\bibnamefont {Sang}},\ }\bibfield  {title} {\enquote
  {\bibinfo {title} {{P-Channel InGaN/GaN heterostructure
  metal-oxide-semiconductor field effect transistor based on
  polarization-induced two-dimensional hole gas}},}\ }\href {\doibase
  10.1038/srep23683} {\bibfield  {journal} {\bibinfo  {journal} {Sci. Rep.}\
  }\textbf {\bibinfo {volume} {6}},\ \bibinfo {pages} {23683} (\bibinfo {year}
  {2016})}\BibitemShut {NoStop}%
\bibitem [{\citenamefont {Nomoto}\ \emph {et~al.}(2017)\citenamefont {Nomoto},
  \citenamefont {Bader}, \citenamefont {Lee}, \citenamefont {Bharadwaj},
  \citenamefont {Hu}, \citenamefont {Xing},\ and\ \citenamefont
  {Jena}}]{Nomoto2017}%
  \BibitemOpen
  \bibfield  {author} {\bibinfo {author} {\bibfnamefont {K.}~\bibnamefont
  {Nomoto}}, \bibinfo {author} {\bibfnamefont {S.~J.}\ \bibnamefont {Bader}},
  \bibinfo {author} {\bibfnamefont {K.}~\bibnamefont {Lee}}, \bibinfo {author}
  {\bibfnamefont {S.}~\bibnamefont {Bharadwaj}}, \bibinfo {author}
  {\bibfnamefont {Z.}~\bibnamefont {Hu}}, \bibinfo {author} {\bibfnamefont
  {H.~G.}\ \bibnamefont {Xing}}, \ and\ \bibinfo {author} {\bibfnamefont
  {D.}~\bibnamefont {Jena}},\ }\bibfield  {title} {\enquote {\bibinfo {title}
  {{Wide-bandgap Gallium Nitride p-channel MISFETs with enhanced performance at
  high temperature}},}\ }in\ \href {\doibase 10.1109/DRC.2017.7999466} {\emph
  {\bibinfo {booktitle} {Device Research Conference - Conference Digest,
  DRC}}}\ (\bibinfo {year} {2017})\BibitemShut {NoStop}%
\bibitem [{\citenamefont {Nakajima}\ \emph {et~al.}(2018)\citenamefont
  {Nakajima}, \citenamefont {Kubota}, \citenamefont {Tsutsui}, \citenamefont
  {Kakushima}, \citenamefont {Wakabayashi}, \citenamefont {Iwai}, \citenamefont
  {Nishizawa},\ and\ \citenamefont {Ohashi}}]{Nakajima2018}%
  \BibitemOpen
  \bibfield  {author} {\bibinfo {author} {\bibfnamefont {A.}~\bibnamefont
  {Nakajima}}, \bibinfo {author} {\bibfnamefont {S.}~\bibnamefont {Kubota}},
  \bibinfo {author} {\bibfnamefont {K.}~\bibnamefont {Tsutsui}}, \bibinfo
  {author} {\bibfnamefont {K.}~\bibnamefont {Kakushima}}, \bibinfo {author}
  {\bibfnamefont {H.}~\bibnamefont {Wakabayashi}}, \bibinfo {author}
  {\bibfnamefont {H.}~\bibnamefont {Iwai}}, \bibinfo {author} {\bibfnamefont
  {S.-i.}\ \bibnamefont {Nishizawa}}, \ and\ \bibinfo {author} {\bibfnamefont
  {H.}~\bibnamefont {Ohashi}},\ }\bibfield  {title} {\enquote {\bibinfo {title}
  {{GaN-based complementary metal-oxide-semiconductor inverter with normally
  off Pch and Nch MOSFETs fabricated using polarisation-induced holes and
  electron channels}},}\ }\href {\doibase 10.1049/iet-pel.2017.0376} {\bibfield
   {journal} {\bibinfo  {journal} {IET Power Elec.}\ }\textbf {\bibinfo
  {volume} {11}},\ \bibinfo {pages} {689} (\bibinfo {year} {2018})}\BibitemShut
  {NoStop}%
\bibitem [{\citenamefont {Krishna}\ \emph {et~al.}(2019)\citenamefont
  {Krishna}, \citenamefont {Raj}, \citenamefont {Hatui}, \citenamefont
  {Romanczyk},\ and\ \citenamefont {Koksaldi}}]{Krishna2019}%
  \BibitemOpen
  \bibfield  {author} {\bibinfo {author} {\bibfnamefont {A.}~\bibnamefont
  {Krishna}}, \bibinfo {author} {\bibfnamefont {A.}~\bibnamefont {Raj}},
  \bibinfo {author} {\bibfnamefont {N.}~\bibnamefont {Hatui}}, \bibinfo
  {author} {\bibfnamefont {B.}~\bibnamefont {Romanczyk}}, \ and\ \bibinfo
  {author} {\bibfnamefont {O.}~\bibnamefont {Koksaldi}},\ }\bibfield  {title}
  {\enquote {\bibinfo {title} {{Gallium nitride ( GaN ) superlattice ( SL )
  based p-channel field effect transistor}},}\ }\href@noop {} {\bibfield
  {journal} {\bibinfo  {journal} {In review}\ ,\ \bibinfo {pages} {1}}
  (\bibinfo {year} {2019})},\ \Eprint {http://arxiv.org/abs/1902.02022}
  {arXiv:1902.02022} \BibitemShut {NoStop}%
\bibitem [{\citenamefont {Ponce}, \citenamefont {Jena},\ and\ \citenamefont
  {Giustino}()}]{Ponce2019}%
  \BibitemOpen
  \bibfield  {author} {\bibinfo {author} {\bibfnamefont {S.}~\bibnamefont
  {Ponce}}, \bibinfo {author} {\bibfnamefont {D.}~\bibnamefont {Jena}}, \ and\
  \bibinfo {author} {\bibfnamefont {F.}~\bibnamefont {Giustino}},\ }\bibfield
  {title} {\enquote {\bibinfo {title} {{Route to high hole mobility in GaN via
  reversal of crystal-field splitting}},}\ }\href@noop {} {\bibinfo  {journal}
  {In review}\ }\BibitemShut {NoStop}%
\bibitem [{\citenamefont {Foreman}(1993)}]{Foreman1993}%
  \BibitemOpen
\bibfield  {journal} {  }\bibfield  {author} {\bibinfo {author} {\bibfnamefont
  {B.~A.}\ \bibnamefont {Foreman}},\ }\bibfield  {title} {\enquote {\bibinfo
  {title} {{Effective-mass Hamiltonian and boundary conditions for the valence
  bands of semiconductor microstructures}},}\ }\href {\doibase
  10.1103/PhysRevB.48.4964} {\bibfield  {journal} {\bibinfo  {journal} {Phys.
  Rev. B}\ }\textbf {\bibinfo {volume} {48}},\ \bibinfo {pages} {4964}
  (\bibinfo {year} {1993})}\BibitemShut {NoStop}%
\bibitem [{\citenamefont {Burt}(1994)}]{Burt1994}%
  \BibitemOpen
  \bibfield  {author} {\bibinfo {author} {\bibfnamefont {M.~G.}\ \bibnamefont
  {Burt}},\ }\bibfield  {title} {\enquote {\bibinfo {title} {{Direct derivation
  of effective-mass equations for microstructures with atomically abrupt
  boundaries}},}\ }\href {\doibase 10.1103/PhysRevB.50.7518} {\bibfield
  {journal} {\bibinfo  {journal} {Phys. Rev. B}\ }\textbf {\bibinfo {volume}
  {50}},\ \bibinfo {pages} {7518} (\bibinfo {year} {1994})}\BibitemShut
  {NoStop}%
\bibitem [{\citenamefont {Mireles}\ and\ \citenamefont
  {Ulloa}(1999)}]{Mireles1999}%
  \BibitemOpen
  \bibfield  {author} {\bibinfo {author} {\bibfnamefont {F.}~\bibnamefont
  {Mireles}}\ and\ \bibinfo {author} {\bibfnamefont {S.}~\bibnamefont
  {Ulloa}},\ }\bibfield  {title} {\enquote {\bibinfo {title} {{Ordered
  hamiltonian and matching conditions for heterojunctions with wurtzite
  symmetry: (formula presented) quantum wells}},}\ }\href {\doibase
  10.1103/PhysRevB.60.13659} {\bibfield  {journal} {\bibinfo  {journal} {Phys.
  Rev. B - Condensed Matter and Materials Physics}\ }\textbf {\bibinfo {volume}
  {60}},\ \bibinfo {pages} {13659} (\bibinfo {year} {1999})}\BibitemShut
  {NoStop}%
\bibitem [{\citenamefont {Chuang}\ and\ \citenamefont
  {Chang}(1996)}]{Chuang1996}%
  \BibitemOpen
  \bibfield  {author} {\bibinfo {author} {\bibfnamefont {S.~L.}\ \bibnamefont
  {Chuang}}\ and\ \bibinfo {author} {\bibfnamefont {C.~S.}\ \bibnamefont
  {Chang}},\ }\bibfield  {title} {\enquote {\bibinfo {title} {{K⋅P Method for
  Strained Wurtzite Semiconductors}},}\ }\href {\doibase
  10.1103/PhysRevB.54.2491} {\bibfield  {journal} {\bibinfo  {journal} {Phys.
  Rev. B}\ }\textbf {\bibinfo {volume} {54}},\ \bibinfo {pages} {2491 -- 2504}
  (\bibinfo {year} {1996})}\BibitemShut {NoStop}%
\bibitem [{\citenamefont {Birner}(2011)}]{BirnerThesis}%
  \BibitemOpen
  \bibfield  {author} {\bibinfo {author} {\bibfnamefont {S.}~\bibnamefont
  {Birner}},\ }\emph {\bibinfo {title} {{Modeling of semiconductor
  nanostructures and semiconductor – electrolyte interfaces}}},\ \href@noop
  {} {Ph.D. thesis},\ \bibinfo  {school} {Technical University of Munich}
  (\bibinfo {year} {2011})\BibitemShut {NoStop}%
\bibitem [{\citenamefont {Bader}()}]{PyNitride}%
  \BibitemOpen
  \bibfield  {author} {\bibinfo {author} {\bibfnamefont {S.~J.}\ \bibnamefont
  {Bader}},\ }\href@noop {} {\enquote {\bibinfo {title} {{PyNitride}},}\
  }\bibinfo {howpublished} {\url{http://sambader.net/pynitride}}\BibitemShut
  {NoStop}%
\bibitem [{\citenamefont {Tan}\ \emph {et~al.}(1990)\citenamefont {Tan},
  \citenamefont {Snider}, \citenamefont {Chang},\ and\ \citenamefont
  {Hu}}]{Tan1990}%
  \BibitemOpen
  \bibfield  {author} {\bibinfo {author} {\bibfnamefont {I.-H.}\ \bibnamefont
  {Tan}}, \bibinfo {author} {\bibfnamefont {G.~L.}\ \bibnamefont {Snider}},
  \bibinfo {author} {\bibfnamefont {L.~D.}\ \bibnamefont {Chang}}, \ and\
  \bibinfo {author} {\bibfnamefont {E.~L.}\ \bibnamefont {Hu}},\ }\bibfield
  {title} {\enquote {\bibinfo {title} {{A self-consistent solution of
  Schrödinger–Poisson equations using a nonuniform mesh}},}\ }\href
  {\doibase 10.1063/1.346245} {\bibfield  {journal} {\bibinfo  {journal} {J.
  App. Phys.}\ }\textbf {\bibinfo {volume} {68}},\ \bibinfo {pages} {4071}
  (\bibinfo {year} {1990})}\BibitemShut {NoStop}%
\bibitem [{\citenamefont {Stroscio}\ and\ \citenamefont
  {Dutta}(2004)}]{StroscioDutta}%
  \BibitemOpen
  \bibfield  {author} {\bibinfo {author} {\bibfnamefont {M.~A.}\ \bibnamefont
  {Stroscio}}\ and\ \bibinfo {author} {\bibfnamefont {M.}~\bibnamefont
  {Dutta}},\ }\href@noop {} {\emph {\bibinfo {title} {{Phonons in
  Nanostructures}}}}\ (\bibinfo  {publisher} {Cambridge University Press},\
  \bibinfo {address} {Cambridge},\ \bibinfo {year} {2004})\BibitemShut
  {NoStop}%
\bibitem [{\citenamefont {Shi}(2003)}]{Shi2003}%
  \BibitemOpen
  \bibfield  {author} {\bibinfo {author} {\bibfnamefont {J.}~\bibnamefont
  {Shi}},\ }\bibfield  {title} {\enquote {\bibinfo {title} {{Interface
  optical-phonon modes and electron–interface-phonon interactions in wurtzite
  GaN/AlN quantum wells}},}\ }\href {\doibase 10.1103/PhysRevB.68.165335}
  {\bibfield  {journal} {\bibinfo  {journal} {Phys. Rev. B}\ }\textbf {\bibinfo
  {volume} {68}},\ \bibinfo {pages} {31} (\bibinfo {year} {2003})}\BibitemShut
  {NoStop}%
\bibitem [{\citenamefont {Medeiros}\ \emph {et~al.}(2005)\citenamefont
  {Medeiros}, \citenamefont {Albuquerque}, \citenamefont {Farias},
  \citenamefont {Vasconcelos},\ and\ \citenamefont {Anselmo}}]{Medeiros2005}%
  \BibitemOpen
  \bibfield  {author} {\bibinfo {author} {\bibfnamefont {S.~K.}\ \bibnamefont
  {Medeiros}}, \bibinfo {author} {\bibfnamefont {E.~L.}\ \bibnamefont
  {Albuquerque}}, \bibinfo {author} {\bibfnamefont {G.~A.}\ \bibnamefont
  {Farias}}, \bibinfo {author} {\bibfnamefont {M.~S.}\ \bibnamefont
  {Vasconcelos}}, \ and\ \bibinfo {author} {\bibfnamefont {D.~H. A.~L.}\
  \bibnamefont {Anselmo}},\ }\bibfield  {title} {\enquote {\bibinfo {title}
  {{Confinement of polar optical phonons in AlN/GaN superlattices}},}\ }\href
  {\doibase 10.1016/j.ssc.2005.02.027} {\bibfield  {journal} {\bibinfo
  {journal} {Solid State Communications}\ }\textbf {\bibinfo {volume} {135}},\
  \bibinfo {pages} {144} (\bibinfo {year} {2005})}\BibitemShut {NoStop}%
\bibitem [{\citenamefont {Liao}, \citenamefont {Dutta},\ and\ \citenamefont
  {Stroscio}(2010)}]{Liao2010}%
  \BibitemOpen
  \bibfield  {author} {\bibinfo {author} {\bibfnamefont {S.}~\bibnamefont
  {Liao}}, \bibinfo {author} {\bibfnamefont {M.}~\bibnamefont {Dutta}}, \ and\
  \bibinfo {author} {\bibfnamefont {M.~A.}\ \bibnamefont {Stroscio}},\
  }\bibfield  {title} {\enquote {\bibinfo {title} {{Interface optical phonon
  modes in wurtzite quantum heterostructures}},}\ }\href {\doibase
  10.1109/IWCE.2010.5677978} {\bibfield  {journal} {\bibinfo  {journal} {Intl.
  Workshop on Computational Electronics}\ }\textbf {\bibinfo {volume}
  {054312}},\ \bibinfo {pages} {215} (\bibinfo {year} {2010})}\BibitemShut
  {NoStop}%
\bibitem [{\citenamefont {Komirenko}\ \emph {et~al.}(1999)\citenamefont
  {Komirenko}, \citenamefont {Kim}, \citenamefont {Stroscio},\ and\
  \citenamefont {Dutta}}]{Komirenko1999a}%
  \BibitemOpen
  \bibfield  {author} {\bibinfo {author} {\bibfnamefont {S.}~\bibnamefont
  {Komirenko}}, \bibinfo {author} {\bibfnamefont {K.}~\bibnamefont {Kim}},
  \bibinfo {author} {\bibfnamefont {M.}~\bibnamefont {Stroscio}}, \ and\
  \bibinfo {author} {\bibfnamefont {M.}~\bibnamefont {Dutta}},\ }\bibfield
  {title} {\enquote {\bibinfo {title} {{Dispersion of polar optical phonons in
  wurtzite quantum wells}},}\ }\href {\doibase 10.1103/PhysRevB.59.5013}
  {\bibfield  {journal} {\bibinfo  {journal} {Phys. Rev. B}\ }\textbf {\bibinfo
  {volume} {59}},\ \bibinfo {pages} {5013} (\bibinfo {year}
  {1999})}\BibitemShut {NoStop}%
\bibitem [{\citenamefont {Zhu}, \citenamefont {Ban},\ and\ \citenamefont
  {Ha}(2012)}]{Zhu2012}%
  \BibitemOpen
  \bibfield  {author} {\bibinfo {author} {\bibfnamefont {J.}~\bibnamefont
  {Zhu}}, \bibinfo {author} {\bibfnamefont {S.~L.}\ \bibnamefont {Ban}}, \ and\
  \bibinfo {author} {\bibfnamefont {S.~H.}\ \bibnamefont {Ha}},\ }\bibfield
  {title} {\enquote {\bibinfo {title} {{Phonon-assisted intersubband
  transitions in wurtzite GaN/InxGa1-xN quantum wells}},}\ }\href {\doibase
  10.1088/1674/21/9/097301} {\bibfield  {journal} {\bibinfo  {journal} {Chinese
  Physics B}\ }\textbf {\bibinfo {volume} {21}},\ \bibinfo {pages} {1}
  (\bibinfo {year} {2012})}\BibitemShut {NoStop}%
\bibitem [{\citenamefont {Pokatilov}, \citenamefont {Nika},\ and\ \citenamefont
  {Balandin}(2003)}]{Pokatilov2003}%
  \BibitemOpen
  \bibfield  {author} {\bibinfo {author} {\bibfnamefont {E.~P.}\ \bibnamefont
  {Pokatilov}}, \bibinfo {author} {\bibfnamefont {D.~L.}\ \bibnamefont {Nika}},
  \ and\ \bibinfo {author} {\bibfnamefont {A.~A.}\ \bibnamefont {Balandin}},\
  }\bibfield  {title} {\enquote {\bibinfo {title} {{Phonon spectrum and group
  velocities in AlN/GaN/AlN and related heterostructures}},}\ }\href {\doibase
  10.1016/S0749(03)00069} {\bibfield  {journal} {\bibinfo  {journal}
  {Superlattices and Microstructures}\ }\textbf {\bibinfo {volume} {33}},\
  \bibinfo {pages} {155} (\bibinfo {year} {2003})}\BibitemShut {NoStop}%
\bibitem [{\citenamefont {Pokatilov}, \citenamefont {Nika},\ and\ \citenamefont
  {Balandin}(2004)}]{Pokatilov2004}%
  \BibitemOpen
  \bibfield  {author} {\bibinfo {author} {\bibfnamefont {E.~P.}\ \bibnamefont
  {Pokatilov}}, \bibinfo {author} {\bibfnamefont {D.~L.}\ \bibnamefont {Nika}},
  \ and\ \bibinfo {author} {\bibfnamefont {A.~A.}\ \bibnamefont {Balandin}},\
  }\bibfield  {title} {\enquote {\bibinfo {title} {{Confined electron-confined
  phonon scattering rates in wurtzite AlN/GaN/AlN heterostructures}},}\ }\href
  {\doibase 10.1063/1.1710705} {\bibfield  {journal} {\bibinfo  {journal} {J.
  App. Phys.}\ }\textbf {\bibinfo {volume} {95}},\ \bibinfo {pages} {5626}
  (\bibinfo {year} {2004})}\BibitemShut {NoStop}%
\bibitem [{\citenamefont {Balandin}, \citenamefont {Nika},\ and\ \citenamefont
  {Pokatilov}(2004)}]{Balandin2004}%
  \BibitemOpen
  \bibfield  {author} {\bibinfo {author} {\bibfnamefont {A.~A.}\ \bibnamefont
  {Balandin}}, \bibinfo {author} {\bibfnamefont {D.~L.}\ \bibnamefont {Nika}},
  \ and\ \bibinfo {author} {\bibfnamefont {E.~P.}\ \bibnamefont {Pokatilov}},\
  }\bibfield  {title} {\enquote {\bibinfo {title} {{Phonon spectrum and group
  velocities in wurtzite hetero-structures}},}\ }\href {\doibase
  10.1002/pssc.200405418} {\bibfield  {journal} {\bibinfo  {journal} {Physica
  Status Solidi C: Conferences}\ }\textbf {\bibinfo {volume} {1}},\ \bibinfo
  {pages} {2658} (\bibinfo {year} {2004})}\BibitemShut {NoStop}%
\bibitem [{\citenamefont {Balandin}, \citenamefont {Pokatilov},\ and\
  \citenamefont {Nika}(2007)}]{Balandin2007}%
  \BibitemOpen
  \bibfield  {author} {\bibinfo {author} {\bibfnamefont {A.~A.}\ \bibnamefont
  {Balandin}}, \bibinfo {author} {\bibfnamefont {E.~P.}\ \bibnamefont
  {Pokatilov}}, \ and\ \bibinfo {author} {\bibfnamefont {D.~L.}\ \bibnamefont
  {Nika}},\ }\bibfield  {title} {\enquote {\bibinfo {title} {{Phonon
  Engineering in Hetero- and Nanostructures}},}\ }\href {\doibase
  10.1166/jno.2007.201} {\bibfield  {journal} {\bibinfo  {journal} {J.
  Nanoelectronics and Optoelectronics}\ }\textbf {\bibinfo {volume} {2}},\
  \bibinfo {pages} {140} (\bibinfo {year} {2007})}\BibitemShut {NoStop}%
\bibitem [{\citenamefont {Vurgaftman}\ and\ \citenamefont
  {Meyer}(2003)}]{VM2003}%
  \BibitemOpen
  \bibfield  {author} {\bibinfo {author} {\bibfnamefont {I.}~\bibnamefont
  {Vurgaftman}}\ and\ \bibinfo {author} {\bibfnamefont {J.~R.}\ \bibnamefont
  {Meyer}},\ }\bibfield  {title} {\enquote {\bibinfo {title} {{Band parameters
  for nitrogen-containing semiconductors}},}\ }\href {\doibase
  10.1063/1.1600519} {\bibfield  {journal} {\bibinfo  {journal} {J. App.
  Phys.}\ }\textbf {\bibinfo {volume} {94}},\ \bibinfo {pages} {3675} (\bibinfo
  {year} {2003})}\BibitemShut {NoStop}%
\bibitem [{\citenamefont {Suzuki}\ and\ \citenamefont
  {Uenoyama}(1996)}]{Suzuki1996}%
  \BibitemOpen
  \bibfield  {author} {\bibinfo {author} {\bibfnamefont {M.}~\bibnamefont
  {Suzuki}}\ and\ \bibinfo {author} {\bibfnamefont {T.}~\bibnamefont
  {Uenoyama}},\ }\bibfield  {title} {\enquote {\bibinfo {title} {{Reduction of
  threshold current density of wurtzite GaN/AlGaN quantum well lasers by
  uniaxial strain in (0001) plane}},}\ }\href@noop {} {\bibfield  {journal}
  {\bibinfo  {journal} {Japanese J. App. Phys.}\ }\textbf {\bibinfo {volume}
  {35}},\ \bibinfo {pages} {L953} (\bibinfo {year} {1996})}\BibitemShut
  {NoStop}%
\bibitem [{\citenamefont {Dasgupta}, \citenamefont {Radosavljevic},\ and\
  \citenamefont {Then}(2017)}]{DasguptaPatent2017}%
  \BibitemOpen
  \bibfield  {author} {\bibinfo {author} {\bibfnamefont {S.}~\bibnamefont
  {Dasgupta}}, \bibinfo {author} {\bibfnamefont {M.}~\bibnamefont
  {Radosavljevic}}, \ and\ \bibinfo {author} {\bibfnamefont {H.~W.}\
  \bibnamefont {Then}},\ }\href@noop {} {\enquote {\bibinfo {title} {{Stressors
  for compressively strained GaN p-channel}},}\ }\bibinfo {howpublished} {WO
  2017/099752 Al 15} (\bibinfo {year} {2017})\BibitemShut {NoStop}%
\bibitem [{\citenamefont {Gupta}\ \emph {et~al.}(2018)\citenamefont {Gupta},
  \citenamefont {Tsukada}, \citenamefont {Romanczyk}, \citenamefont {Pasayat},
  \citenamefont {James}, \citenamefont {Keller},\ and\ \citenamefont
  {Mishra}}]{Gupta2018}%
  \BibitemOpen
  \bibfield  {author} {\bibinfo {author} {\bibfnamefont {C.}~\bibnamefont
  {Gupta}}, \bibinfo {author} {\bibfnamefont {Y.}~\bibnamefont {Tsukada}},
  \bibinfo {author} {\bibfnamefont {B.}~\bibnamefont {Romanczyk}}, \bibinfo
  {author} {\bibfnamefont {S.~S.}\ \bibnamefont {Pasayat}}, \bibinfo {author}
  {\bibfnamefont {D.-a.}\ \bibnamefont {James}}, \bibinfo {author}
  {\bibfnamefont {S.}~\bibnamefont {Keller}}, \ and\ \bibinfo {author}
  {\bibfnamefont {U.~K.}\ \bibnamefont {Mishra}},\ }\bibfield  {title}
  {\enquote {\bibinfo {title} {{First experimental demonstration of enhancement
  in hole conductivity in c-plane (0001) III-Nitrides with uniaxial strain}},}\
  }in\ \href@noop {} {\emph {\bibinfo {booktitle} {Intl. Workshop on
  Nitrides}}}\ (\bibinfo {address} {Kanazawa, Japan},\ \bibinfo {year}
  {2018})\BibitemShut {NoStop}%
\end{thebibliography}%

\end{document}